\documentstyle[12pt]{article}
\input math_macros.tex

\begin{document}
\begin{titlepage}
\begin{center}
Dec. 18,1997     \hfill    LBNL-41188 \\

\vskip .5in

{\large \bf Reply to Unruh}
\footnote{This work was supported by the Director, Office of Energy 
Research, Office of High Energy and Nuclear Physics, Division of High 
Energy Physics of the U.S. Department of Energy under Contract 
DE-AC03-76SF00098.}
\vskip .50in
Henry P. Stapp\\
{\em Lawrence Berkeley National Laboratory\\
      University of California\\
    Berkeley, California 94720}
\end{center}

\vskip .5in

\begin{abstract}

William Unruh has suggested (quant-ph/9710032) that a certain counterfactual
statement in my recent nonlocality proof should be re-interpreted in a way 
that would block the proof. I give reason's why that statement should not
be re-interpreted.  

\end{abstract}
\medskip
\end{titlepage}

\renewcommand{\thepage}{\roman{page}}
\setcounter{page}{2}
\mbox{ }

\vskip 1in

\begin{center}
{\bf Disclaimer}
\end{center}

\vskip .2in

\begin{scriptsize}
\begin{quotation}
This document was prepared as an account of work sponsored by the United
States Government. While this document is believed to contain correct 
 information, neither the United States Government nor any agency
thereof, nor The Regents of the University of California, nor any of their
employees, makes any warranty, express or implied, or assumes any legal
liability or responsibility for the accuracy, completeness, or usefulness
of any information, apparatus, product, or process disclosed, or represents
that its use would not infringe privately owned rights.  Reference herein
to any specific commercial products process, or service by its trade name,
trademark, manufacturer, or otherwise, does not necessarily constitute or
imply its endorsement, recommendation, or favoring by the United States
Government or any agency thereof, or The Regents of the University of
California.  The views and opinions of authors expressed herein do not
necessarily state or reflect those of the United States Government or any
agency thereof or The Regents of the University of California and shall
not be used for advertising or product endorsement purposes.
\end{quotation}
\end{scriptsize}

\vskip 2in

\begin{center}
\begin{small}
{\it Lawrence Berkeley Laboratory is an equal opportunity employer.}
\end{small}
\end{center}

\newpage
\renewcommand{\thepage}{\arabic{page}}
\setcounter{page}{1}

Unruh$^1$ emphasizes that, in the use counterfactuals within a quantum 
context, great care is required to ensure that there is no importation of
classical notions of reality.

Great care is certainly required in both directions: we must neither
allow improper importations of classical notions of reality, nor blind 
ourselves to unexpected properties of nature by placing arbitrary constraints
on rational analysis. 

After all, the basic difficulty that caused the founders of 
quantum theory to insist that the quantum formalism had to be interpreted 
as being about ``our knowledge'', instead of being about physical
reality itself---as physical reality had formerly been understood in 
physics---was the need to reduce  wave packets of large extent upon the 
receipt of new information. This reduction is natural for ``our 
knowledge'', but conflicts with locality ideas about physical reality coming 
from (deterministic relativistic) classical physics. 

In view of the controversial nature of this Copenhagen move---of making
physics be about human knowledge---it is certainly proper to question 
whether this peculiar way of evading a nonlocality that is so blatantly 
present in the mathematical formalism might not be obscuring a nonlocal aspect 
that is actually present in nature herself. So we must be as much on guard 
against curtailing rational argumentation as against importing  classical
ideas about reality: this issue is too important to be settled by classically 
based prejudices of any kind.

The entire argument is about macroscopic events, such as the setting up of 
alternative possible experiments, and constraints on the possibilities for 
outcomes of such experiments. At that level Bohr advocated the use of 
classical language and logic, and emphasized the freedom of experimenters to 
examine properties of their own  choosing. The entire Bohr-EPR discussion was 
based on the common agreement that consideration of mutually exclusive 
alternative possible measurements was not out of bounds. 

The step in my proof$^2$ that Unruh objects to is the step where LOC2 is 
applied. I had shown that under the condition that L2 is performed a 
certain statement (S) is true:

(S): If the first measurement was performed in region R and gave the 
first of the two possible outcomes there then if, instead, the second 
possible measurement had been performed there then the outcome would have 
been the first possible outcome of that second measurement. 

I take this statement (S) to mean what it says: there is some deep structure
in nature that connects what actually occurs under the actually realized
experimental conditions to what would have occurred if the quantum/free 
choice that determined which experiment was performed in region R had 
gone the other way.

This property was proved, without appeal to determinism or hidden variables,
under the condition that L2 was performed (plus the condition LOC1,
and the assumed validity of the predictions of quantum theory in this 
Hardy-type case, and the idea that the choices as to which experiments
are performed in the two regions can be treated as independent free variables) 

Because experiment L2 is supposed to be performed after everything 
referred to in (S) has either occurred or not, I claim that if (S) should 
{\it fail} to hold under the condition that the later free choice were  L1, 
instead of L2, then there must be some sort of backward-in-time influence: 
the constraints connecting possibilities of outcomes in region R that is 
asserted to hold by statement (S) would either hold or not hold according to 
whether or not the later free choice is to perform L2 or L1. LOC2 is, 
accordingly, the postulate that (S) {\it continues} to hold if L1 is performed
at the later time, rather than L2. [Since the conjunction of the postulates 
leads to a contradiction, this postulate is, to be sure, a very likely 
candidate for rejection.]

Unruh claims that a hidden classical reality assumption is smuggled in here.

I have always stressed in my work on this subject that I am excluding  
the many-worlds scenarios, in which nature makes no choices: the entire 
argument is based on the notion of the lack of dependence of nature's choice
of which outcome appears upon which free choice is made (later) 
by a faraway experimenter. This is a reality assumption that I do make.

Most quantum theorist, when not adhering strictly to the Copenhagen
position that the theory is about our knowledge, do think in these terms:
nature selects the outcomes of the quantum measurements that we choose to 
perform. This is not a classical idea of reality, because it is about
a stochastic selection that has no counterpart in classical mechanics.
It is a quantum idea about reality.

Unruh argues that, because a certain outcome of L2 occurs in the 
{\it proof} of (S) under condition L2, a failure of property (S) 
to remain true if L1 is performed, instead of L2, would not constitute
a backward in time influence. For him the meaning of (S) is entangled 
with its proof. I, on the other hand, adhere consistently throughout my 
proof to the position the (S) is defined as a condition that might or 
might not be true: it asserts that nature has an aspect that 
connects what actually happens if one of the experiments is performed
in region R to what would have happened there if the quantum event that
controls which measurement is performed there had gone the other way.
I show that such a constraint can in fact be proved to exist under the 
condition that L2 is performed. 

This interpretation of (S) is its natural meaning: it is what the words say.
In any case, it is completely rational for me to consistently interpret (S) 
this way. For this is the interpretation that, consistently applied,
allows us to explore most effectively the character of a possible quantum 
reality that lies behind the phenomena that is the source of our knowledge of
nature.

{\bf References}

1. W. Unruh, {\it Is Quantum Mechanics Non-local?} (quant-ph/9710032)

2. H.P. Stapp. {\it Nonlocal Character of Quantum Theory}, Amer.
J. Phys. {\bf 65}, 300-304 (1997) 
(http://www-physics.lbl.gov/~stapp/stappfiles.html)

\end{document}